\documentclass[showpacs,nofootinbib,preprintnumbers,prd,superscriptaddress,twocolumn]{revtex4-1}

\usepackage{amsmath}
\usepackage{amsfonts}
\usepackage{amssymb}
\usepackage{graphicx}
\usepackage[titletoc]{appendix}
\usepackage{color}
\usepackage{hyperref}
\usepackage{cleveref}
\usepackage[rightcaption]{sidecap}
\usepackage{comment}
\usepackage{bm}
\usepackage{subfig}
\usepackage{physics}
\usepackage{dsfont}
\usepackage[normalem]{ulem}
\usepackage{mathrsfs}
\usepackage{calrsfs}
\usepackage{bbold}
\usepackage{mathtools}
\usepackage{dcolumn}
\usepackage{array}
\usepackage{ctable}
\usepackage{multirow}
\usepackage{physics}
\usepackage{tabularx}
\usepackage{booktabs}

\DeclareMathAlphabet{\mathcal}{OMS}{cmsy}{m}{n}

\graphicspath{{Graphics/}}
\def\be{\begin{equation}}
\def\ee{\end{equation}}
\def\bea{\begin{eqnarray}}
\def\eea{\end{eqnarray}}

\definecolor{vividviolet}{rgb}{0.62, 0.0, 1.0}
\definecolor{amaranth}{rgb}{0.9, 0.17, 0.31}
\definecolor{palatinateblue}{rgb}{0.15, 0.23, 0.89}
\definecolor{brightpink}{rgb}{1.0, 0.0, 0.5}
\definecolor{cornflowerblue}{rgb}{0.39, 0.58, 0.93}
\definecolor{deepcarminepink}{rgb}{0.94, 0.19, 0.22}
\definecolor{radicalred}{rgb}{1.0, 0.21, 0.37}

\hypersetup{ linktoc=all,
    colorlinks, linkcolor={palatinateblue},
    citecolor={brightpink}, urlcolor={amaranth}
}

\begin{document}

\title{Horizon entanglement area law from regular black hole thermodynamics}

\author{Alessio Belfiglio}
\email{alessio.belfiglio@unicam.it}
\affiliation{School of Science and Technology, University of Camerino, Via Madonna delle Carceri, Camerino, 62032, Italy.}
\affiliation{Istituto Nazionale di Fisica Nucleare (INFN), Sezione di Perugia, Perugia, 06123, Italy.}

\author{S. Mahesh Chandran}
\email{maheshchandran@iitb.ac.in}
\affiliation{Department of Physics, Indian Institute of Technology Bombay, Mumbai 400076, India.}

\author{Orlando Luongo}
\email{orlando.luongo@unicam.it}
\affiliation{School of Science and Technology, University of Camerino, Via Madonna delle Carceri, Camerino, 62032, Italy.}
\affiliation{Istituto Nazionale di Fisica Nucleare (INFN), Sezione di Perugia, Perugia, 06123, Italy.}
\affiliation{SUNY Polytechnic Institute, 13502 Utica, New York, USA.}
\affiliation{INAF - Osservatorio Astronomico di Brera, Milano, Italy.}
\affiliation{Al-Farabi Kazakh National University, Al-Farabi av. 71, 050040 Almaty, Kazakhstan.}

\author{Stefano Mancini}
\email{stefano.mancini@unicam.it}
\affiliation{School of Science and Technology, University of Camerino, Via Madonna delle Carceri, Camerino, 62032, Italy.}
\affiliation{Istituto Nazionale di Fisica Nucleare (INFN), Sezione di Perugia, Perugia, 06123, Italy.}

\begin{abstract}
We investigate the thermodynamics of regular black hole configurations via quantum analogs of entropy and energy --- namely, the entanglement entropy and entanglement energy --- near the event horizon of Bardeen and Hayward black holes. Following standard approaches, we introduce a quantum scalar field propagating in such black hole spacetimes and discretize the field degrees of freedom on a lattice of spherical shells. We observe that, at leading order, the entanglement entropy associated with the scalar field is proportional to Bekestein-Hawking entropy, while the corresponding entanglement energy scales proportionally to Komar energy. We then compute the heat capacity in both scenarios, discussing the black hole stability conditions and the possible appearance of second-order phase transitions. Finally, we extend our analysis to the black hole core, showing that in this sector entanglement energy serves as a valuable tool towards discriminating between singular and regular solutions.
\end{abstract}


\maketitle


\section{Introduction} \label{intro}

The role of entanglement entropy in characterizing quantum systems has been widely investigated in recent years, with important applications spanning from quantum information processing \cite{PhysRevLett.70.1895,PhysRevA.53.2046, Bennett2000}, to condensed
matter physics \cite{PhysRevLett.90.227902, PhysRevLett.93.260602, PhysRevLett.94.050501, RicoOrtega2022} and quantum field theory \cite{PhysRevD.34.373,PhysRevLett.71.666, PasqualeCalabrese_2004}.

One usually expects that the entropy associated with a given system in a distinguished region should exhibit an extensive character, resulting in a volume scaling, which is indeed peculiar to thermal states. However, the situation is generally different for ground states of quantum systems, which typically fulfill an \emph{area law}, with small logarithmic corrections in some cases \cite{RevModPhys.82.277}. The emergence of an area law is intimately connected to the presence of local interactions within quantum many-body systems, i.e., the quantum correlations between a given region and its exterior are typically dominated by the short-range couplings across the boundary surface \cite{PhysRevLett.90.227902, Hastings_2007, Brandão2013}.

For example, Ref. \cite{PhysRevD.34.373} showed the existence of an area law associated with a free, massless scalar field in $(3+1)$ flat spacetime, tracing over the (discretized) degrees of freedom inside a distinguished region and computing the von Neumann entropy \cite{RevModPhys.81.865, Wilde_2017} corresponding to the reduced density operator for this state. A similar result was later obtained in the case of a spherical entangling surface \cite{PhysRevLett.71.666}.

Motivated by these studies, discrete field theories have attracted significant interest in recent years, focusing in particular on possible connections with the widely-studied Bekenstein-Hawking entropy for black holes \cite{PhysRevD.7.2333,1975HawkingCommun.Math.Phys.,doi:10.1080/00107510310001632523}. Despite a complete understanding of black hole entropy would ultimately require a consistent theory of quantum gravity\footnote{See, e.g., \cite{BarberoG:2022ixy} and references therein for some proposals in the framework of Loop Quantum Gravity.}, recent progress suggests that entanglement entropy serves as a fundamental tool in this direction \cite{VPFrolov_1998, Jacobson2003,Das2010, Casini_2009, Nishioka_2009, Calabrese_2009, Solodukhin2011,2020Chandran.ShankaranarayananPhys.Rev.D}. Specifically, the discretization scheme proposed in \cite{PhysRevLett.71.666} was later generalized to the case of static, spherically symmetric configurations \cite{PhysRevD.77.064013}, confirming that the ground state entanglement entropy of a minimally coupled scalar field remains consistent with an area law behaviour\footnote{Power-law corrections may arise when the field is in an excited state, but fall off with increasing area, thereby recovering the area law for large horizons.}. More recently, scalar field ground state entropy was also studied in the context of Schwarzschild-like quantum black hole spacetimes \cite{Belfiglio:2024qsa}, suggesting possible deviations from area law at Planck scales.

The presence of non-minimal coupling to the background scalar curvature can also play a key role in entanglement generation, leading to area law violations for sufficiently large coupling constants  \cite{BELFIGLIO2024138398}. This outcome may have relevant consequences in the strong gravity regime, particularly in the early-universe scenarios pertaining to primordial black holes \cite{10.1093/mnras/168.2.399,PaulH.Frampton_2010, PhysRevLett.117.061101} and inflation \cite{Tsujikawa:2003jp,RevModPhys.78.537, Baumann:2009ds}, where entanglement characterization is gaining considerable attention, see e.g. \cite{Arias_2020, PhysRevD.102.043529, PhysRevD.107.103512, PhysRevD.108.043522,2024Chandran.etalPhys.Rev.D}.

In the absence of a complete quantum gravity theory \cite{10.1093/acprof:oso/9780199585205.001.0001, Carney_2019}, entanglement characterization in curved spacetimes then emerges as a remarkable approach that encompasses fundamental aspects of quantum gravity, without being contingent upon specific details of any underlying theory. In this respect, the thermodynamics of black holes may provide hints toward the quantum nature of these objects \cite{PhysRevD.13.191, Wald:1999vt}, shedding light on the origin of black hole entropy and the corresponding horizon properties \cite{Das2010}. While entanglement entropy and black-hole entropy are intimately connected via the area-law, a full thermodynamic picture requires suitable quantum counterparts for energy and temperature to also be defined, that can potentially satisfy a quantum analog of the first-law. To this end, various definitions of \textit{entanglement energy} and \textit{entanglement temperature} were previously studied in order to construct a thermodynamic picture arising from quantum measures (referred to as entanglement thermodynamics) and understand its connection with the thermodynamics of the background geometry~\cite{1997-Mukohyama.etal-Phys.Rev.D,Mukohyama1997,1998-Mukohyama.etal-Phys.Rev.D}. Such a construction has important implications for spacetimes with horizons, which represent physical boundaries beyond which the observers do not have access to information. That is, while in flat backgrounds the chosen boundary for area law calculations is merely artificial, entanglement characterization close to a spacetime horizon captures relevant information about its thermodynamic structure, thus suggesting that horizon thermodynamics may in itself be of quantum origin \cite{2020Chandran.ShankaranarayananPhys.Rev.D}.

Motivated by the above results, in this work we address the issue of black hole horizon stability by considering the corresponding entanglement thermodynamics, focusing in particular on regular configurations. Very recently the interest in regular black holes has increased, after the imaging of accretion disks around black holes \cite{Akiyama_2019e1,Akiyama_2019e2,Akiyama_2019e3,Akiyama_2019e4,Akiyama_2019e5}. The physics of a regular black hole is analogous to a standard black hole, exhibiting the same thermodynamics \cite{PhysRevD.97.104015, Huang_2024}, albeit the existence of singularity is avoided by an extra term, i.e. topological charge, vacuum energy, etc. Recently, the Penrose theorems \cite{PhysRevLett.14.57, Hawking:1970zqf} related to the existence of singularities have been critically revised by Kerr \cite{Kerr:2023rpn}, emphasizing that regular objects may in principle be more present than expected \footnote{Even the physics of compact objects has been revised through regular configurations, with particular regards to quasi-periodic oscillations \cite{Bambi:2008hp, PhysRevD.108.044063}, showing that such solutions may be in principle adaptable to astrophysical objects in the universe.}. Exploiting the near-horizon approximation \cite{PhysRevD.71.104029, PhysRevD.71.124040}, we compute the entanglement entropy and the corresponding entanglement energy for Bardeen \cite{1968qtr..conf...87B} and Hayward \cite{PhysRevLett.96.031103} regular black holes, showing how the Smarr formula for black hole thermodynamics can be recovered in the continuum limit. We then address the issue of thermodynamic black hole stability at the horizon, computing the heat capacity in both scenarios and observing the emergence of second-order phase transitions.  Finally, we extend our analysis to the core of such black hole configurations, \emph{demonstrating that entanglement energy may allow to discriminate between regular and singular solutions at sufficiently small length-scales}. These outcomes suggest that entanglement characterization can play a key role in the attempt to understand the quantum nature of black hole thermodynamics.

The paper is organized as follows. In Sec. \ref{sec:warmup}, we review the scalar field discretization procedure in static and spherically symmetric spacetimes, tracing back the field degrees of freedom to a system of coupled harmonic oscillators. In Sec. \ref{sec:setup}, we introduce the notions of entanglement energy and entanglement entropy for discrete field theories in the near-horizon approximation. In Sec. \ref{sec:nearhorizon}, we study the entanglement thermodynamics and horizon stability of Bardeen and Hayward black holes. In Sec. \ref{sec:core}, we move inside the black hole core and highlight the main differences between regular and singular solutions in terms of entanglement energy. Finally, in Sec. \ref{sec:conc} we discuss our outcomes and draw our conclusions\footnote{Throughout this work, we adopt the metric signature $(-,+,+, +)$ and set $G=\hbar=c=k_B=4\pi\epsilon_0=1$.}.

\section{Geometric warm up}\label{sec:warmup}

We start with a static, spherically symmetric spacetime, described by the line-element
\begin{equation}
	\label{eq:linelemrad}
	ds^2=-f(r)dt^2+\frac{1}{f(r)}dr^2+r^2d\Omega^2,
\end{equation}
written in the Schwarzschild coordinates. Then, introducing the proper-length,
\begin{equation}\label{eq:properlength}
	\rho=\int_{r_h}^r \frac{dr'}{\sqrt{f(r')}},
\end{equation}
it is convenient to rewrite Eq. \eqref{eq:linelemrad} in terms of $\rho$ as coordinate, useful for spacetimes having horizons \cite{1998-Mukohyama.etal-Phys.Rev.D}. Thus, we obtain


\begin{equation}
	\label{eq:linelemprop}
	ds^2=-f(r)dt^2+d\rho^2+r^2d\Omega^2.
\end{equation}

The selected initial value $r_h$, in Eq. \eqref{eq:properlength}, represents the horizon radius, satisfying $f(r_h)=0$, and corresponds to $\rho=0$.

Specifically, it is worth noticing that the proper length coordinates can only describe the geometry in the region $f(r) > 0$ from the horizon and, so, if the horizon vanishes, namely $r_h\to 0$, the proper length reduces to the radial coordinate, i.e. $\rho\to r$.

\subsection{Scalar field dynamics}

We now consider a weak scalar field $\varphi$ propagating in the classical background geometry described by Eq. \eqref{eq:linelemprop}, with corresponding action~\cite{1982-Birrell.Davies-QuantumFieldsCurved,Jacobson2005}

\begin{eqnarray}\label{eq:action}
	\mathcal S=\frac{1}{2}\int d^4x\sqrt{-g}\left[g^{\mu\nu}\partial_{\mu}\varphi\partial_{\nu}\varphi-m^2\varphi^2\right],
\end{eqnarray}

where $m$ is the particle mass. Exploiting spherical symmetry, we can perform the usual partial wave expansion of the scalar field~\cite{PhysRevLett.71.666,2020Chandran.ShankaranarayananPhys.Rev.D}, thus writing
\begin{subequations} \label{sphericaldec}
    \begin{align}
	\dot{\varphi}(\rho,\theta,\phi)&=\frac{f^{1/4}}{r}\sum_{lm}\dot{\varphi}_{lm}(\rho)Z_{lm}(\theta,\phi),\\
	\varphi(\rho,\theta,\phi)&=\frac{f^{1/4}}{r}\sum_{lm}\varphi_{lm}(\rho)Z_{lm}(\theta,\phi).
    \end{align}
\end{subequations}
The above expansion leads to the following effective (1+1)-D field Lagrangian ~\cite{1997-Mukohyama.etal-Phys.Rev.D,1998-Mukohyama.etal-Phys.Rev.D}
	\begin{multline}\label{eq:lag}
		\mathcal L=\frac{1}{2}\sum_{lm}\int d\rho\Bigg[\dot{\varphi}_{lm}^2-r^2(\rho)\sqrt{f}\bigg\{\partial_{\rho}\bigg(\frac{f^{1/4}\varphi_{lm}}{r(\rho)}\bigg)\bigg\}^2\\-f\left\{m_f^2+\frac{l(l+1)}{r^2(\rho)}\right\}\varphi_{lm}^2\Bigg].
\end{multline}
To quantize the field, we promote the conjugate variables, $\varphi_{lm}$ and $\pi_{lm} \equiv \dot{\varphi}_{lm}$, to operators satisfying the usual commutation relations,
\begin{subequations}
    \begin{align} &[\hat{\varphi}_{lm},\hat{\pi}_{l'm'}]=i\delta_{lm}^{l'm'}\,,\\
&[\hat{\varphi}_{lm},\hat{\varphi}_{l'm'}]=[\hat{\pi}_{lm},\hat{\pi}_{l'm'}]=0.
    \end{align}
\end{subequations}

We now exploit the standard procedure of ultraviolet (UV) regularization by discretizing the scalar field degrees of freedom via a lattice of spherical shells, with spacing $a$. The discretized proper length coordinate then reads
\begin{equation}
\rho=ja,
\end{equation}
and the radial distance between consequent spherical shells introduces a UV cutoff, given by $1/a$. Similarly, the overall size of the lattice
\begin{equation}
\rho_L=(N+1)a.
\end{equation}
imposes the corresponding IR cutoff, namely $1/\rho_L$. For each lattice point $j$ along proper length, we obtain the corresponding lattice point along radial coordinate as $r_j=a^{-1}r(\rho)|_{\rho=ja}$, where $r(\rho)$ is obtained by inverting the expression for proper length in Eq. \eqref{eq:linelemprop}.

Hence, in the limit $m\rightarrow0$, we end up with the following Hamiltonian
\begin{widetext}	
	\begin{equation}\label{eq:Hamil}
		H=\frac{1}{2a}\sum_{lmj}\left[\pi_{lm,j}^2+r_{j+\frac{1}{2}}^2f_{j+\frac{1}{2}}^{1/2}\bigg\{\frac{f_j^{1/4}\varphi_{lm,j}}{r_j}-\frac{f_{j+1}^{1/4}\varphi_{lm,j+1}}{r_{j+1}}\bigg\}^2+\frac{f_jl(l+1)}{r_j^2}\varphi_{lm,j}^2\right],
	\end{equation}
\end{widetext}
where $f_j=f\left[r(\rho=ja)\right]$. Eq. \eqref{eq:Hamil} describes a \emph{regularized massless quantum field propagating in a classical background spacetime}, while the flat spacetime Hamiltonian is recovered in the limit $f_j \rightarrow 1$ ~\cite{2020Chandran.ShankaranarayananPhys.Rev.D}.

We notice that Eq. \eqref{eq:Hamil} can be traced back to a system of coupled harmonic oscillators, writing
\begin{equation}\label{eq:hlm}
	H_{lm}=\frac{1}{2}\left[\sum_i \pi_{lm,i}^2+\sum_{ij}K_{ij}\varphi_{lm,i}\varphi_{lm,j}\right],
\end{equation}
and, then, decomposing Eq. \eqref{eq:Hamil} by
\begin{equation}\label{eq:lmHamil}
H=\frac{1}{a}\sum_{lm} H_{lm},
\end{equation}
with $K_{ij}$ the coupling matrix for each $lm$-lattice. Upon imposing Dirichlet boundary conditions, $\varphi(0)=\varphi(\rho_L)=0$, the coupling matrix acquires the following non-zero elements,
\begin{small}
\begin{subequations}
     \begin{align}\label{eq:Kmatrix}
		K_{jj}=&\frac{f_jl(l+1)}{r_j^2}+\frac{\sqrt{f_j}}{r_j^2}\left[r_{j+\frac{1}{2}}^2\sqrt{f_{j+\frac{1}{2}}}+r_{j-\frac{1}{2}}^2\sqrt{f_{j-\frac{1}{2}}}\right],\\
		K_{j,j+1}=&K_{j+1,j}=-\frac{f_{j+\frac{1}{2}}^{1/2}f_{j}^{1/4}f_{j+1}^{1/4}r_{j+\frac{1}{2}}^2}{r_jr_{j+1}}.
	\end{align}
\end{subequations}
\end{small}

\section{Near-Horizon Approximation}\label{sec:setup}

Close to the horizons, the expression in Eq. \eqref{eq:Hamil} can be further simplified. In this limit, we can approximate the lapse function by
\begin{equation}\label{eq:lapseapprox}
f(r)\approx (r-r_h)f'(r_h),
\end{equation}
where it may be noted that $f^\prime(r_h)$ is related to the Hawking temperature as $T_H=f'(r_h)/4\pi$~\cite{1973-Bardeen.etal-CMP}. This approximation gives us a Hamiltonian fully-described by just two dimensionless parameters, namely $r_h/a$, $\propto S_{BH}^{1/2}$, and $af'(r_h)$, $\propto T_H$, that are intimately related to the black hole thermodynamics\footnote{Indeed, this near-horizon feature is universal and appears independent of the black hole spacetime we handle.}, as we will clarify later in the text. Eq. \eqref{eq:lapseapprox} then gives
\begin{equation}
\rho\approx 2\sqrt{\frac{r-r_h}{f'(r_h)}},
\end{equation}
implying immediately the following relations:
\begin{subequations}\label{eq:nhrelation}
\begin{align}
r_j&=\frac{r_h}{a}+\frac{j^2af'(r_h)}{4},\\ f_j&=\frac{j^2a^2\left(f'(r_h)\right)^2}{4}.
\end{align}
\end{subequations}
Plugging these relations into Eq.  \eqref{eq:Hamil}, we infer the Hamiltonian description very close to the horizon.  Substituting these relations into Eq. \eqref{eq:Kmatrix} also fixes the coupling matrix for the field degrees of freedom near the horizon.

\subsection{Entanglement thermodynamics at the horizon}\label{ssec:entthermo}

To quantify the true degrees of freedom that give rise to black hole entropy, we can compute the field entanglement across its horizon~\cite{2007DasClassicalandQuantumGravity,2020Chandran.ShankaranarayananPhys.Rev.D}.

As previously stated, in the proper length coordinates we only have access to the regions where $f(r) > 0$. Accordingly, bipartition of the field exactly at the horizon is not possible. However, it is feasible very close to the horizon ~\cite{1997-Mukohyama.etal-Phys.Rev.D,1998-Mukohyama.etal-Phys.Rev.D,2020Chandran.ShankaranarayananPhys.Rev.D}, i.e., we can look at a single-oscillator subsystem in the lattice described in Eq. \eqref{eq:Hamil}, exploiting the near-horizon parameters of Eqs. \eqref{eq:nhrelation}.

Upon spatially partitioning the field into ``in" and ``out" degrees of freedom for each $lm$-mode, the Hamiltonian in Eq. \eqref{eq:hlm} can be split into subsystems of $n$ and $N-n$ oscillators, respectively
\begin{equation}
    H=H_{in}+H_{out}+H_{int},
\end{equation}
with
\begin{subequations} \label{hamsplit}
    \begin{align}
    H_{in}&=\frac{1}{2}\left[\sum_{i=1}^n \pi_i^2+\sum_{ij=1}^nK_{_{ij}}\varphi_{i}\varphi_{j}\right],\\
    H_{out}&=\frac{1}{2}\left[\sum_{i=n+1}^N \pi_{i}^2+\sum_{ij=n+1}^N K_{_{ij}}\varphi_{i}\varphi_{j}\right],\\ H_{int}&=K_{_{n,n+1}}\varphi_{{n}}\varphi_{{n+1}},
    \end{align}
\end{subequations}

\noindent where we have dropped the $lm$-indices from each Hamiltonian in Eq. \eqref{hamsplit} to simplify the notation.

We now study subsystem thermodynamics of the ``in" degrees of freedom with the help of the following measures:
\begin{itemize}
    \item[-]  \textbf{The von-Neumann entropy}. It measures the ``mixedness" of the density matrix describing the ``in"-subsystem:
    \begin{equation}
        S=-\Tr[\rho_{in}\ln\rho_{in}]\,;\quad \rho_{in}=\Tr_{out}\ket{\Psi}\bra{\Psi}
    \end{equation}
    where reduced density matrix $\rho_{in}$ is obtained by tracing away the ``out" degrees of freedom from the total density matrix $\rho=\ket{\Psi}\bra{\Psi}$. If the overall state $\ket{\Psi}$ is pure, then von-Neumann entropy of $\rho_{in}$ serves as a measure of entanglement between ``in" and ``out" degrees of freedom, i.e, it is referred to as \textit{entanglement entropy}.
    \item[-] \textbf{The subsystem energy}. Expectation value of the normal-ordered Hamiltonian corresponding to the ``in"-subsystem:
    \begin{equation}
    E=\langle :H_{in}: \rangle_{in}=\Tr\left[\rho_{in}:H_{in}:\right],
\end{equation}
where,
\begin{multline}\label{eq:hin}
    :H_{in}:\, =\frac{1}{2}\sum_{i=1}^n\left[\pi_i+i\sum_{j=1}^n\left(K_{in}^{1/2}\right)_{ij}\varphi_j\right]\\\cross\left[\pi_i-i\sum_{j=1}^n\left(K_{in}^{1/2}\right)_{ij}\varphi_j\right].
\end{multline}
Here, $K_{in}$ is the $n\cross n$ sub-block of coupling matrix $K$ in Eq. \eqref{eq:Kmatrix} corresponding to only the ``in" degrees of freedom. While this serves as a measure for the disturbed vacuum energy of the ``in" subsystem, it is also one of the few candidates for \textit{entanglement energy} that were previously discussed in literature as suitable quantum analogs for energy, that could potentially reproduce the first law in tandem with the entanglement entropy~\cite{1997-Mukohyama.etal-Phys.Rev.D,1998-Mukohyama.etal-Phys.Rev.D,Mukohyama1997}. For this reason, we will continue to refer to this measure of subsystem energy as entanglement energy for the rest of the paper.
\end{itemize}
Assuming a vacuum configuration for the field, we can describe the $N$-oscillator system using a Gaussian wave-function. In this case, the covariance matrix $\Sigma$ is defined as follows~\cite{serafini2017quantum,2021Jain.etalPhys.Rev.D}:
\begin{equation}\label{eq:cov}
    \Sigma=\begin{bmatrix}       \Sigma_{\varphi\varphi}&\Sigma_{\varphi\pi}\\\Sigma_{\varphi\pi}^T&\Sigma_{\pi\pi}    \end{bmatrix}\,,
\end{equation}
where the block matrices in $\Sigma$ are constructed with two-point correlation functions that contain full information about a Gaussian state:
\begin{subequations}\label{eq:corr}
    \begin{align}
        (\Sigma_{\varphi\varphi})_{ij}&=\frac{1}{2}\langle \{\varphi_i,\varphi_j\}\rangle\\
        (\Sigma_{\varphi\pi})_{ij}&=\frac{1}{2}\langle \{\varphi_i,\pi_j\}\rangle\\
        (\Sigma_{\pi\pi})_{ij}&=\frac{1}{2}\langle \{\pi_i,\pi_j\}\rangle.
    \end{align}
\end{subequations}

Accordingly, given a $n-$oscillator ``in"-subsystem, the reduced covariance matrix, $\sigma_{in}$, is simply obtained from Eq. \eqref{eq:cov} by taking the $n\cross n$ sub-blocks of the $N\cross N$ matrices in Eqs. \eqref{eq:corr}
\begin{subequations}\label{eq:corr2}
    \begin{align}
\sigma_{in}&=\begin{bmatrix}       \sigma_{\varphi\varphi}&\sigma_{\varphi\pi}\\\sigma_{\varphi\pi}^T&\sigma_{\pi\pi}    \end{bmatrix}\,,\\
\left(\sigma_{\varphi\varphi}\right)_{ij}&=\left(\Sigma_{\varphi\varphi}\right)_{ij\in \left\{1,..,n\right\}\,,}\\
\left(\sigma_{\pi\pi}\right)_{ij}&=\left(\Sigma_{\pi\pi}\right)_{ij\in \left\{1,..,n\right\}\,,}\\
\left(\sigma_{\varphi\pi}\right)_{ij}&=\left(\Sigma_{\varphi\pi}\right)_{ij\in \left\{1,..,n\right\}\,.}
    \end{align}
\end{subequations}

The commutation relations for conjugate variables $\{\varphi_i\}=\{\Xi_i\}$ and $\{\pi_i\}=\{\Xi_{i+N}\}$ can be represented as follows:
\begin{equation}
    [\Xi_i,\Xi_j]=i\Omega_{ij}\,,\quad \Omega=\begin{bmatrix}
        O&\mathcal{I}\\-\mathcal{I}&O
    \end{bmatrix}\,.
\end{equation}
These relations are always preserved via the symplectic transformations, $M$, satisfying $M\Omega M^T=\Omega$. The covariance matrix $\Sigma$ can be brought to the Williamson normal form~\cite{serafini2017quantum}, namely
\begin{equation}
    \tilde{\Sigma}=M\Sigma M^T=\begin{bmatrix}
        diag(\gamma_k)&O\\O&diag(\gamma_k)
    \end{bmatrix}\,.
\end{equation}
The symplectic spectrum $\{\gamma_k\}$ can be obtained from the eigenvalues $\{\pm\gamma_k\}$ of the matrix $i\Omega\Sigma$.

Thus, the \textit{entanglement entropy} finally reads
\begin{multline}\label{eq:EntNHO}
    S=\sum_{k=1}^n\Bigg[\left(\gamma_k+\frac{1}{2}\right)\log\left(\gamma_k+\frac{1}{2}\right)-\\\left(\gamma_k-\frac{1}{2}\right)\log\left(\gamma_k-\frac{1}{2}\right)\Bigg],
\end{multline}
where $\{\gamma_k\}$ are the symplectic eigenvalues of $\sigma_{in}$. For time-independent lattices, since $\sigma_{\varphi\pi}=0$, the symplectic eigenvalues reduce to:
\begin{equation}
    \{\gamma_k\}=eig\left\{\sqrt{\sigma_{\varphi\varphi}\sigma_{\pi\pi}}\right\}.
\end{equation}

On the other hand, the \textit{entanglement energy} can be obtained from Eq. \eqref{eq:hin} and Eqs. \eqref{eq:corr2} as follows:
\begin{equation}
    E=\frac{1}{2}\Tr\left[\sigma_{\pi\pi}+K_{in}\sigma_{\varphi\varphi}-K_{in}^{1/2}\right].
\end{equation}

In order to study the entanglement thermodynamics of the field very close to the horizon, we consider the near-horizon Hamiltonian described by Eq. \eqref{eq:Hamil} with the prescriptions of Eq. \eqref{eq:nhrelation}, and we set $n=1$ corresponding to a spherical shell at a very small distance, say $\rho\sim a$, from the horizon.

Consequently, the total entanglement energy and entropy are obtained by summing up spherical symmetry-resolved contributions that converge for very large $l$-values~\cite{PhysRevLett.71.666,1997-Mukohyama.etal-Phys.Rev.D,2020Chandran.ShankaranarayananPhys.Rev.D}:
\begin{equation}
    S=\sum_{lm}(2l+1)S_{lm},\quad aE=\sum_{lm}(2l+1)E_{lm}.
\end{equation}

\section{Thermodynamics and horizon stability}\label{sec:nearhorizon}

Before moving to the computation of entanglement entropy for regular black holes, we briefly summarize the basic demands of black hole thermodynamics.

Precisely, we start with the Smarr formula, namely the black hole analog of the Gibbs-Duhem relation \cite{Smarr:1972kt},
\begin{equation}
	\label{eq:Smarr}
    M=2 T_{H} \, S_{\rm BH} + 2 \Omega_{H} J+\Phi_{H} Q,
\end{equation}
where $M$ is the black hole mass, $T_H$ the Hawking temperature on the horizon, $S_{BH}$ the Bekenstein-Hawking entropy, $\Omega_H$ the angular velocity, $J$ the angular momentum, $\Phi_H$ the electrostatic potential and, finally, $Q$ the black hole charge. The above form serves as a thermodynamic equation of state for the black hole horizon and satisfies the Penrose no-hair theorem \cite{PhysRevLett.14.57}.

An infinitesimal change of mass leads to the first thermodynamic law of black holes~\cite{1973-Bardeen.etal-CMP,1978-Davies-ReportsonProgressinPhysics,2001-Wald-LivingReviewsinRelativity}
\begin{equation}
    dM=T_HdS_{BH}+\Omega_HdJ+\Phi_H dQ.
\end{equation}
We can further address the thermodynamic stability of the horizon by studying the heat capacity\footnote{Interestingly, heat capacity is usually defined with respect to either charge $Q$ or angular momentum $J$, whichever is kept constant. This is the strategy followed to explore a phase transition in Reisser-Nordstr\"om black holes (RNBHs), due to a divergent discontinuity in $C_Q$, see e.g.~\cite{1978-Davies-ReportsonProgressinPhysics,1977HutMonthlyNoticesoftheRoyalAstronomicalSociety}.}
\begin{equation}
    C_H=\left.\frac{\partial M}{\partial T_H}\right\vert_{Q,J}=\left.T_H\frac{\partial S_{BH}}{\partial T_H}\right\vert_{Q,J},
\end{equation}
wherein $C_H<0$ implies instability. For the Schwarzschild solution, this specific heat is always negative, implying that the black hole is unstable when in contact with a thermal bath. This is in contrast with the \textit{mechanical} stability of an isolated Schwarzschild black hole.

For the given functional form of $f(r)$, we see that the Bekenstein-Hawking entropy and the Hawking temperature can be obtained as:
\begin{equation}
    S_{BH}=\pi r_h^2\,,\quad T_{H}=\frac{f'(r_h)}{4\pi}.
\end{equation}
Interestingly, as remarked below Eq. \eqref{eq:lapseapprox}, the Hamiltonian for near-horizon degrees of freedom is fully described by the parameters $a^{-1}r_h$ ($\propto S_{BH}^{1/2}$) and $af'(r_h)$ ($\propto T_H$). Therefore, the quantum field measures of entanglement entropy and subsystem energy near the horizon can only depend on combinations of UV cutoff $a$, Bekenstein-Hawking entropy $S_{BH}$ and Hawking temperature $T_H$, regardless of the spacetime in question. In the next subsection, we proceed to study the exact nature of this dependence for regular black holes, and explore what it can tell us about the thermodynamic stability of the horizon. For further details on the thermodynamics that hereafter we use, refer to Appendix A below.

\subsection{Horizon Thermodynamics of Bardeen black holes}

Bardeen black holes with mass $M$ and charge $q$ can be described by a lapse function of the form:
\begin{equation}\label{eq:bardeen}
    f(r)=1-\frac{2Mr^2}{(r^2+q^2)^{3/2}}.
\end{equation}
We notice that, in general, there is no singularity at $r=0$ when $q\neq 0$. In the limit $q\to0$, we recover the Schwarzschild solution, which is singular at $r=0$. The charge $q$ therefore introduces regularity in the black hole spacetime.

The above metric allows two horizons $r_\pm$ ($r_+>r_-$), that correspond to real, positive roots of $f(r)$:
\begin{align}
    r_+&=\frac{2M}{3}\sqrt{\left\{1+2\cos{\frac{\theta}{3}}\right\}^2-\frac{9q^2}{4M^2}},\\
    r_-&=\frac{2M}{3}\sqrt{\left\{1+2\cos{\left(\frac{\theta-2\pi}{3}\right)}\right\}^2-\frac{9q^2}{4M^2}},\nonumber\\
\end{align}
where we set
\begin{equation} \label{bardangle}
    \theta=\cos^{-1}\left(1-\frac{27q^2}{8M^2}\right).
\end{equation}
The extremal limit $r_+\to r_-$ bounds the charge $q$ from above, namely
\begin{equation}
    0\leq q \leq \frac{4M}{3\sqrt{3}}\sim0.77M\,,
\end{equation}
beyond which there can no longer be a horizon, i.e., $f(r)$ does not have positive roots, and the Smarr formula in Eq. \eqref{eq:Smarr} is no longer applicable. Further, we also see that at the horizon,
\begin{equation}
    f'(r_\pm)=\frac{2Mr_\pm(r_\pm^2-2q^2)}{(r_\pm^2+q^2)^{5/2}}=\frac{1}{r_\pm}\left(1-\frac{3q^2}{r_\pm^2+q^2}\right).
\end{equation}
We therefore have all the ingredients needed to write the Hamiltonian, $\eqref{eq:Hamil}$, for the near-horizon field degrees of freedom at both $r_+$ and $r_-$. From Eqs. \eqref{eq:nhrelation},
\begin{subequations}
\begin{align}\label{eq:nhbardeen}
r_j^{(\pm)}&=\frac{r_\pm}{a}+\frac{j^2a}{4r_\pm}\left(1-\frac{3q^2}{r_\pm^2+q^2}\right),\\
f_j^{(\pm)}&=\frac{j^2a^2}{4r_\pm^2}\left(1-\frac{3q^2}{r_\pm^2+q^2}\right)^2.
\end{align}
\end{subequations}

\begin{figure*}[!ht]
	\begin{center}
		\subfloat[\label{Bardeen1a}][]{%
			\includegraphics[width=0.4\textwidth]{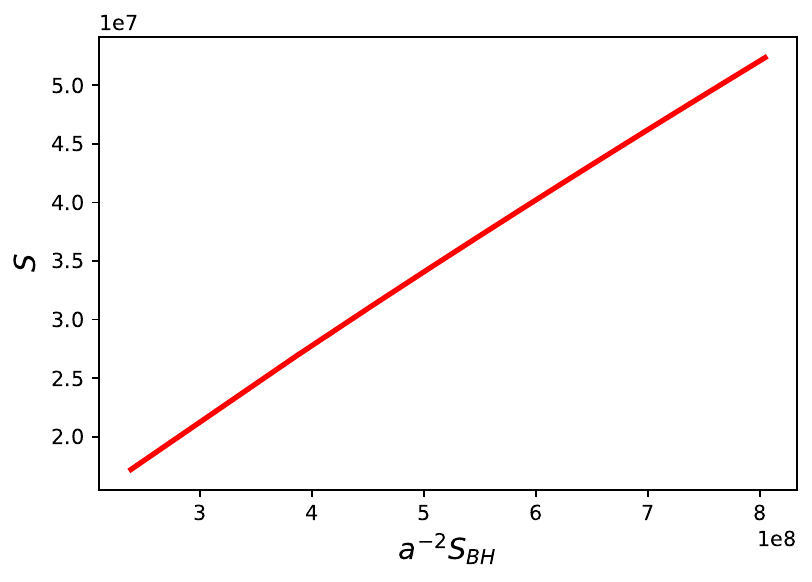}
		}
		\subfloat[\label{Bardeen1b}][]{%
			\includegraphics[width=0.4\textwidth]{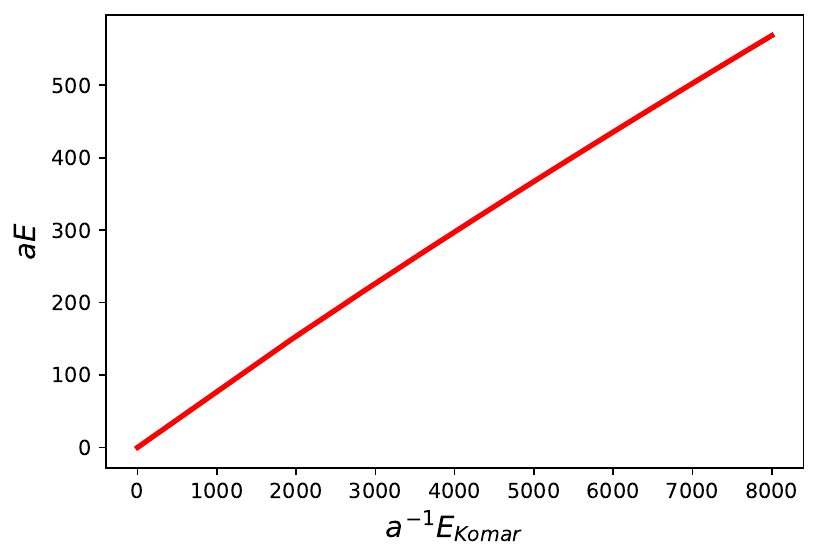}
		}
		
		\caption{Scaling of (a) entanglement entropy with Bekenstein-Hawking entropy, and (b) entanglement energy with Komar energy at proper length $\rho\sim a$ from the event horizon $r_+$ of the Bardeen black hole. Here, $a^{-1}M=8\times10^3$, $N=30$, and we vary the parameter $a^{-1}q$ such that $0< \frac{q}{M} < \frac{4}{3\sqrt{3}}$.}
		\label{fig:Bardeen1}
	\end{center}
\end{figure*}

\begin{figure*}[!ht]
	\begin{center}
		\subfloat[\label{Bardeen2a}][]{%
			\includegraphics[width=0.4\textwidth]{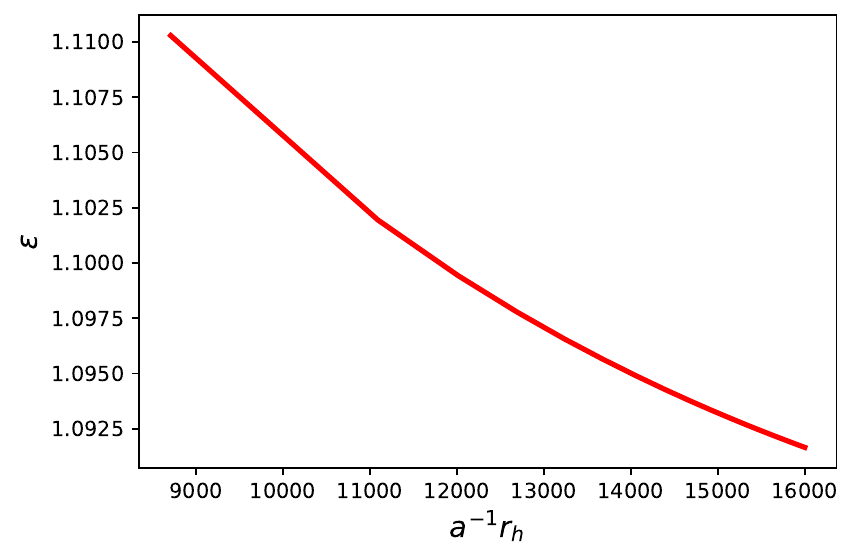}
		}
		\subfloat[\label{Bardeen2b}][]{%
			\includegraphics[width=0.4\textwidth]{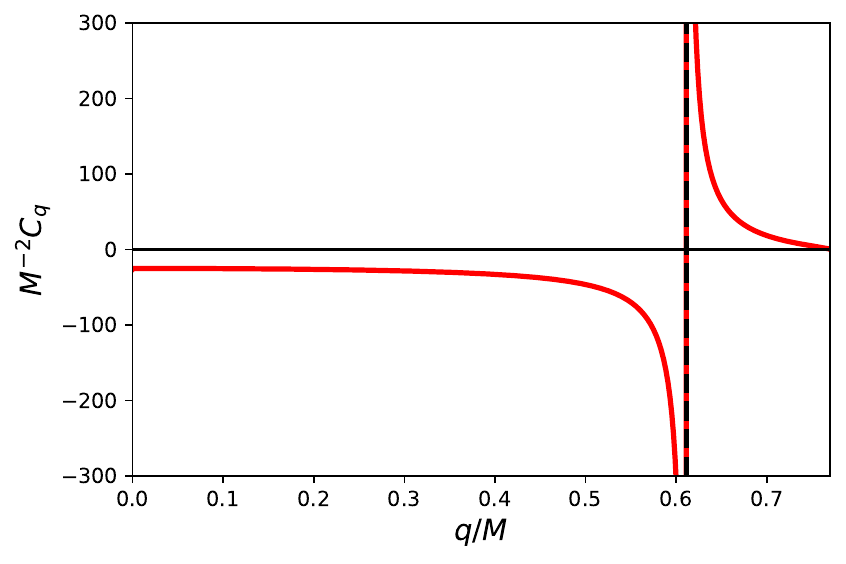}
		}
		
		\caption{(a) The deviation $\epsilon=E/2T_HS$ from Komar relation with respect to horizon radius $r_+$, and (b) the heat capacity per squared mass $M^{-2}C_q$ with respect to the dimensionless parameter $q/M$ for Bardeen black holes.}
		\label{fig:Bardeen2}
	\end{center}
\end{figure*}

Let us now study the thermodynamics of a single-oscillator subsystem very close to the horizon $r_+$, using the tools developed in Sec. \ref{ssec:entthermo}. In this respect, from Fig. \ref{fig:Bardeen1}, we observe the following scaling relations for the entanglement entropy and energy,
\begin{subequations}
    \begin{align}
    &E=\frac{c_e}{a^2}E_{Komar}\,, \label{eq:enescal}\\
    &S=\frac{c_s}{a^2}S_{BH}\,, \label{eq:entroscal}
    \end{align}
\end{subequations}
where
\begin{subequations}
    \begin{align}
        &E_{Komar}=\frac{r_+}{2}\left(1-\frac{3q^2}{r_+^2+q^2}\right),\\
        &S_{BH}=\pi r_+^2,
    \end{align}
\end{subequations}
and $c_e$, $c_s$ are constants, while the Komar energy, $E_{Komar}$, is the conserved charge along the time-like Killing vector defined at the event horizon ~\cite{1959KomarPhys.Rev.,2009-Kastor.etal-CQG,2010-Banerjee.Majhi-Phys.Rev.D,2014-Padmanabhan-GeneralRelativityandGravitation}.

It is therefore easy to see that:
\begin{equation}
    \frac{E}{2S}=\frac{\epsilon E_{Komar}}{2S_{BH}}=\epsilon T_H\,,\quad \epsilon=\frac{c_e}{2c_s},
\end{equation}
where the Hawking temperature $T_H$ is given as
\begin{equation}
    T_H=\frac{f'(r_+)}{4\pi}=\frac{1}{4\pi r_+}\left(1-\frac{3q^2}{r_+^2+q^2}\right).
\end{equation}
Remarkably, we can now stress that $\epsilon$ captures the deviation from the Komar relation, $E_{Komar}=2T_HS_{BH}$. It is possible that this deviation arises from sub-leading corrections or spurious edge effects at the horizon, or from considering the bipartition to be slightly away ($\rho\sim a$) from the horizon. From Fig. \ref{fig:Bardeen2}, we observe that for increasing values of $a^{-1}r_h$,  corresponding to reducing lattice-spacing relative to horizon radius, the deviation $\epsilon$ appears to be monotonically decreasing. Since we ideally require $a\to0$ to recover the continuous field theory from the lattice along with the absence of finite-$N$ effects at the boundary, we expect $\epsilon\to 1$ up to the leading order in this limit. However, a numerical verification of the same is computationally challenging at the present.

This allows us to write down the first law as follows~\cite{2009-Kastor.etal-CQG}:
\begin{equation}
    dM=T_HdS_{BH}+\Phi dq.
\end{equation}
The local thermodynamic stability at the horizon can be studied via heat capacity:
\begin{equation}
    C_q=\left.\frac{\partial M}{\partial T_H}\right\vert_{q}.
\end{equation}
For the Bardeen black hole, we observe that $M^{-2}C_q$ is a function of the ratio $q/M$ alone. From Fig. \ref{fig:Bardeen2}, we see that a second-order phase-transition occurs at $q_c\sim 0.611M$.

Accordingly, the critical value $q_c$ delineates the boundary between the thermodynamically unstable ($C_q<0$) and stable ($C_q>0$) regimes of Bardeen black holes.

\subsection{Horizon Thermodynamics of Hayward Black Holes}

\begin{figure*}[!ht]
	\begin{center}
		\subfloat[\label{Hayward1a}][]{%
			\includegraphics[width=0.4\textwidth]{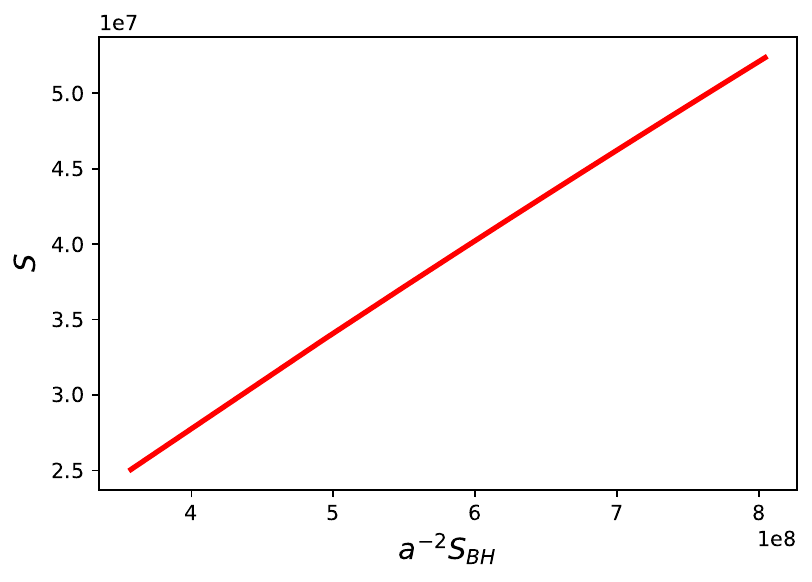}
		}
		\subfloat[\label{Hayward1b}][]{%
			\includegraphics[width=0.4\textwidth]{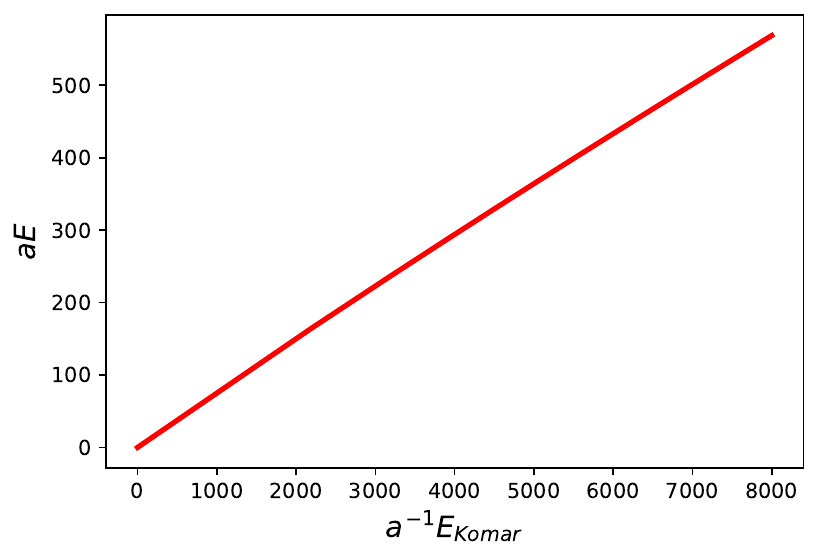}
		}
		
		\caption{Scaling of (a) entanglement entropy with Bekenstein-Hawking entropy, and (b) entanglement energy with Komar energy at proper length $\rho\sim a$ from the event horizon $r_+$ of the Hayward black hole. Here, $a^{-1}M=8\times10^3$ and $N=30$, and we vary the parameter $a^{-1}l$ such that $0< \frac{l}{M} < \frac{4}{3\sqrt{3}}$.}
		\label{fig:Hayward1}
	\end{center}
\end{figure*}

\begin{figure*}[!ht]
	\begin{center}
		\subfloat[\label{Hayward2a}][]{%
			\includegraphics[width=0.4\textwidth]{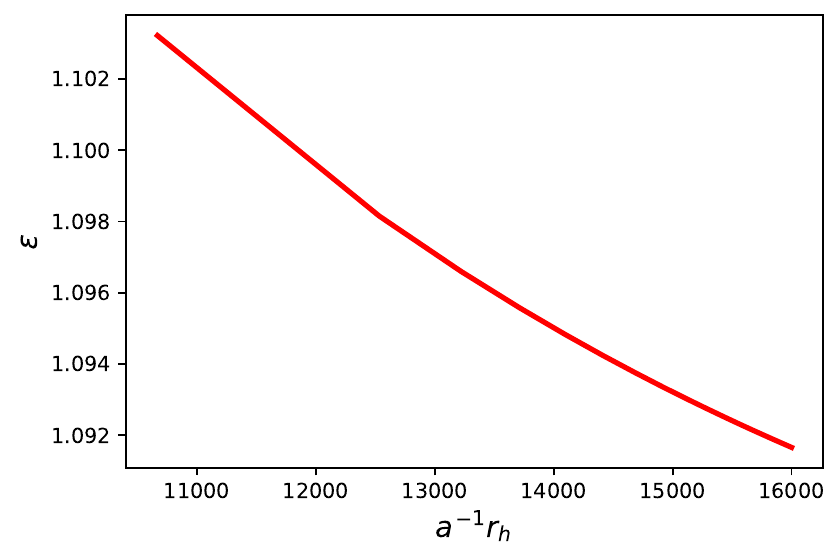}
		}
		\subfloat[\label{Hayward2b}][]{%
			\includegraphics[width=0.4\textwidth]{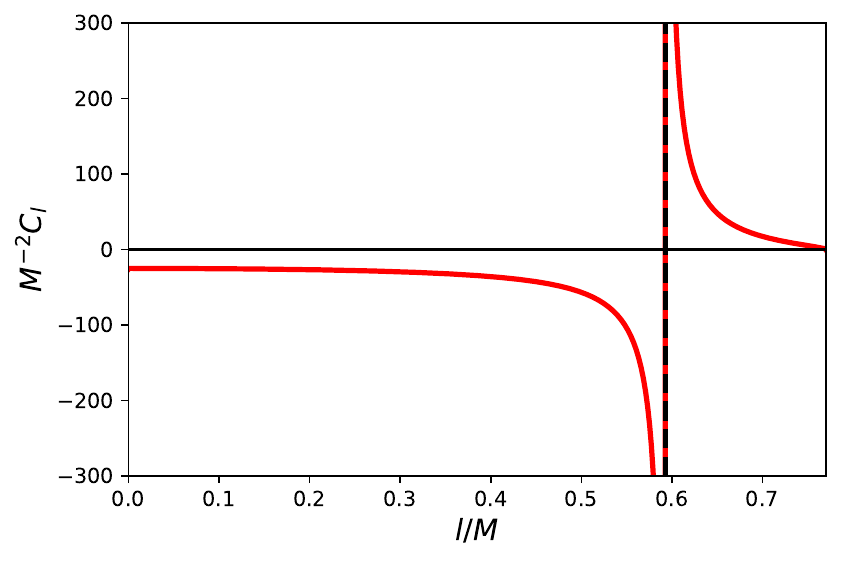}
		}
		
		\caption{(a) The deviation $\epsilon=E/2T_HS$ from Komar relation with respect to horizon radius $r_+$, and (b) the heat capacity per squared mass $M^{-2}C_l$ with respect to the dimensionless parameter $l/M$ for Hayward black holes.}
		\label{fig:Hayward2}
	\end{center}
\end{figure*}

Hayward black holes with mass $M$ and length scale $l$ associated with the regularity at $r=0$ are described by the lapse function
\begin{equation}
    f(r)=1-\frac{2Mr^2}{r^3+2Ml^2}.
\end{equation}
The above metric allows two horizons $r_\pm$ ($r_+>r_-$) again corresponding to real, positive roots of $f(r)$:
\begin{subequations}
    \begin{align}
    r_+&=\frac{2M}{3}\left\{1+2\cos{\frac{\theta}{3}}\right\},\\
    r_-&=\frac{2M}{3}\left\{1+2\cos{\left(\frac{\theta-2\pi}{3}\right)}\right\},\\
    \end{align}
\end{subequations}
where the parameter, $\theta$:
\begin{equation} \label{hayangle}
    \theta=\cos^{-1}\left(1-\frac{27l^2}{8M^2}\right)
\end{equation}
has been introduced, in analogy with Eq. \eqref{bardangle}.
The extremal limit $r_+\to r_-$ bounds the charge $l$ from above:
\begin{equation}
    0\leq l \leq \frac{4M}{3\sqrt{3}}\sim0.77M\,,
\end{equation}
beyond which there is no longer a black hole horizon, i.e., $f(r)>0$ everywhere. We also see that at the horizon:
\begin{equation}
    f'(r_\pm)=\frac{2Mr_\pm(r_\pm^3-4Ml^2)}{(r_\pm^3+2Ml^2)^{2}}=\frac{1}{r_\pm}\left(1-\frac{3l^2}{r_\pm^2}\right).
\end{equation}
Exploiting again Eqs. \eqref{eq:nhrelation}, the near-horizon approximation of \eqref{eq:Hamil} at both $r_+$ and $r_-$ gives now
\begin{subequations}
\begin{align}\label{eq:nhhaywar}
r_j^{(\pm)}&=\frac{r_\pm}{a}+\frac{j^2a}{4r_\pm}\left(1-\frac{3l^2}{r_\pm^2}\right),\nonumber\\
f_j^{(\pm)}&=\frac{j^2a^2}{4r_\pm^2}\left(1-\frac{3l^2}{r_\pm^2}\right)^2.
\end{align}
\end{subequations}
From Fig. \ref{fig:Hayward1}, we notice that in Hayward spacetime the scaling of entanglement entropy and energy is consistent with Eqs. \eqref{eq:enescal}-\eqref{eq:entroscal}, where now
\begin{subequations}
    \begin{align}
       & E_{Komar}=\frac{r_+}{2}\left(1-\frac{3l^2}{r_+^2}\right)
       & S_{BH}=\pi r_+^2
    \end{align}
\end{subequations}
Therefore, we can write again
\begin{equation}
    \frac{E}{2S}=\frac{\epsilon E_{Komar}}{2S_{BH}}=\epsilon T_H,\quad \epsilon=\frac{c_e}{2c_s},
\end{equation}
where the Hawking temperature $T_H$ is now given by
\begin{equation}
    T_H=\frac{f'(r_+)}{4\pi}=\frac{1}{4\pi r_+}\left(1-\frac{3l^2}{r_+^2}\right).
\end{equation}
Similar to the case of Bardeen black holes, $\epsilon$ captures the deviation from the Komar relation as $E_{Komar}=2\epsilon T_HS_{BH}$. From Fig. \ref{fig:Hayward2}, we see that for increasing values of $a^{-1}r_h$, the deviation $\epsilon$ appears to be monotonically decreasing. Since the continuum limit corresponds to infinitely large $a^{-1}r_h$, and upon removing the finite-$N$ effects at the boundary, we again expect $\epsilon\to 1$ up to the leading order in this limit. However, similar to the Bardeen case, a numerical verification of the same is computationally challenging at the present.

This allows us to write down the first law as follows,
\begin{equation}
    dM=T_HdS_{BH}+\Phi dl.
\end{equation}
and to study again the thermodynamic stability at the horizon via the heat capacity, which now reads
\begin{equation}
    C_l=\left.\frac{\partial M}{\partial T_H}\right\vert_{l}.
\end{equation}
For the Hayward black hole, we observe that $M^{-2}C_l$ is a function of the ratio $l/M$ alone.

From Fig. \ref{fig:Hayward2}, we see that a second-order phase-transition occurs at $l_c\sim 0.593M$. The critical value $l_c$ delineates the boundary between the thermodynamically unstable ($C_l<0$) and stable ($C_l>0$) regimes of Hayward black holes.

\section{Entanglement Thermodynamics near the black hole core}\label{sec:core}

The aim of this section is to work out the entanglement thermodynamics near $r=0$. We employ the same Hamiltonian as in \eqref{eq:Hamil}, where the proper length is measured from the (inner) horizon as follows:
\begin{equation}\label{eq:properlength2}
\rho=\int_{r}^{r_h} \frac{dr'}{\sqrt{f(r')}},\quad \rho_L=\rho(r=0)
\end{equation}
Here, we have a natural IR cutoff\footnote{To better clarify its use, for a fixed UV-cutoff $a$, the IR regulator $L=(N+1)a$ is an artificial boundary as it is impractical to simulate the field over the entire space with $N\to\infty$ oscillators. There is also an implicit assumption here that there are no non-trivial effects coming from the field at spatial infinities. In the case of $r<r_-$, there is a natural boundary at $r=0$ in terms of proper length from the inner horizon as $\rho_L=(N+1)a$. This ensures that the entire profiles of entanglement entropy and energy inside the core can be captured using finite values of $N$.} at $\rho_L=(N+1)a$, corresponding to the center $r=0$, thereby ensuring that the Hamiltonian in Eq. \eqref{eq:Hamil} remains Hermitian, i.e., $f_j>0$. Similar to the approach taken in previous sections, we fix the bipartition in a way that allows us to study the entanglement thermodynamics of the subsystem near $r=0$, i.e., we place the boundary at $n=N-1$ and study the subsystem measures corresponding to the single-oscillator ``out" subsystem ($K_{out}$ here being the smallest field sub-region closest to $r=0$) by varying the parameters of the black hole spacetime in question.

Since the effects of singularity/regularity on entanglement mechanics can be better observed in the region $0\leq r\leq r_-$, we now look into the cores of Bardeen, Hayward, and Reissner Nordstr\"om black holes.

\begin{figure*}[!ht]
	\begin{center}
		\subfloat[\label{dscore1}][]{%
			\includegraphics[width=0.4\textwidth]{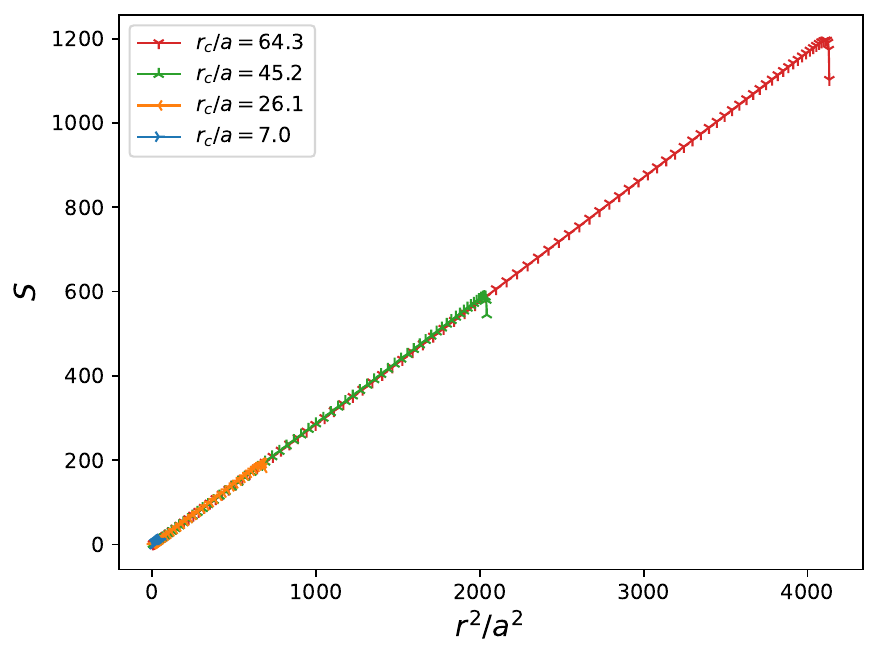}
		}
		\subfloat[\label{dscore2}][]{%
			\includegraphics[width=0.4\textwidth]{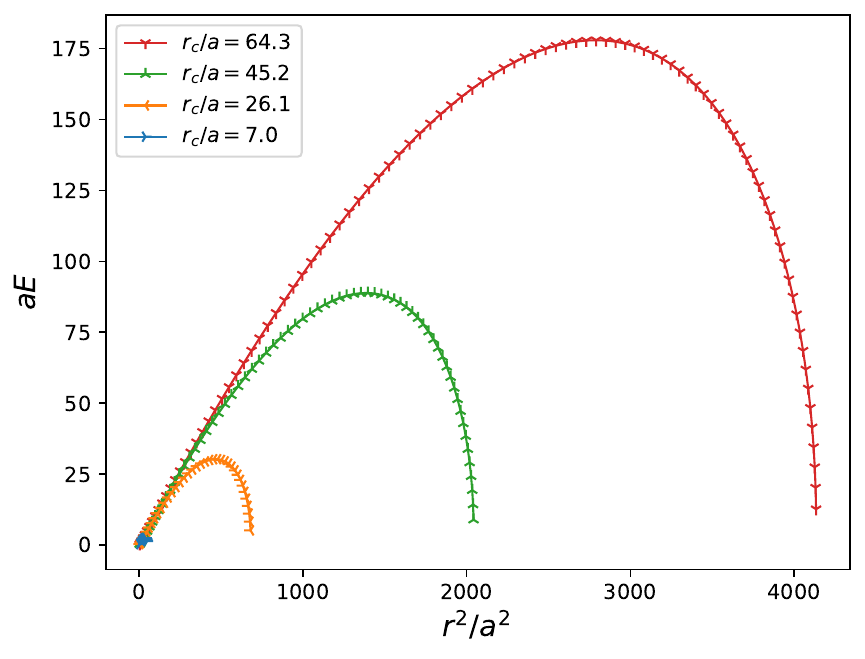}
		}
		
		\caption{Subsystem scaling of (a) entanglement entropy and (b) entanglement energy for different values of rescaled radius ($r_c/a$) for the de Sitter core. Here, the system size ($N+1=\rho_L/a$) is naturally fixed by the proper length to the centre $r=0$, corresponding to each value of $r_c/a$.}
		\label{fig:dScore}
	\end{center}
\end{figure*}

\begin{figure*}[!ht]
	\begin{center}
		\subfloat[\label{rnbh1}][]{%
			\includegraphics[width=0.4\textwidth]{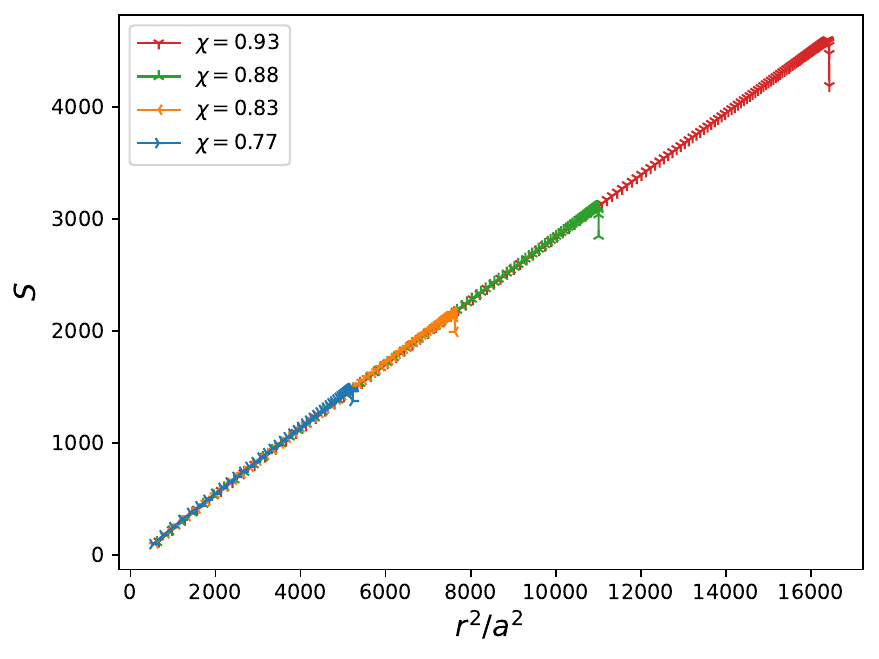}
		}
		\subfloat[\label{rnbh2}][]{%
			\includegraphics[width=0.4\textwidth]{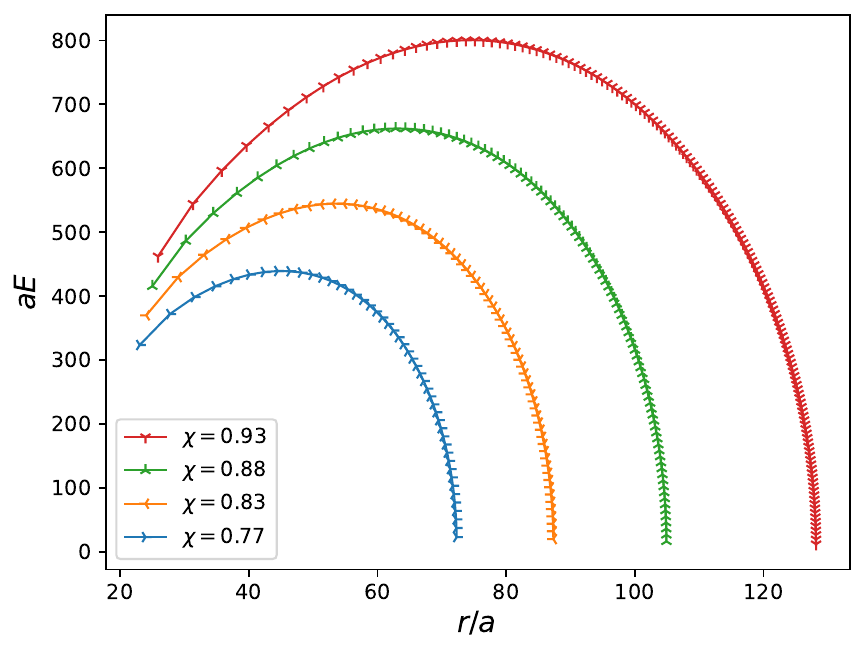}
		}
		
		\caption{Subsystem scaling of (a) entanglement entropy and (b) entanglement energy inside the Cauchy horizon of Reissner-Nordstr\"om black hole for different $\chi=Q/M$ values, with $M/a=100$. Here, the system size ($N+1=\rho_L/a$) is naturally fixed by the proper length to the centre $r=0$, corresponding to each value of $\chi$. Note that the lattice points in radial coordinates are however noticeably different for each choice of $\chi$.}
		\label{fig:RNBHCore}
	\end{center}
\end{figure*}

\subsection{de Sitter core}
In the limit $r\to0$, the Bardeen and Hayward spacetimes resemble a de Sitter core:
\begin{eqnarray}
    f(r)\sim 1-\frac{r^2}{r_c^2},\quad r_c = \begin{cases}
        \frac{q^{3/2}}{\sqrt{2M}} & \text{Bardeen,}\\ l & \text{Hayward.}
    \end{cases}
\end{eqnarray}
The proper length from horizon for de-Sitter core is obtained below~\cite{2020Chandran.ShankaranarayananPhys.Rev.D}:
\begin{equation}
    \rho=r_c \cos ^{-1}\left(\frac{r}{r_c}\right)
\end{equation}
The IR cutoff is therefore fixed corresponding to $r=0$ as:
\begin{equation}
    \rho_L=(N+1)a=\frac{\pi r_c}{2}
\end{equation}
From Fig. \ref{fig:dScore}, we observe that the entanglement entropy scaling follows an area-law throughout, whereas entanglement energy progressively appears to converge to an area-law only as $r\to0$. This is consistent with the fact that for $r\ll r_c$, the spacetime is Minkowski-flat ($f(r)\to 1$), due to which entanglement entropy and energy are expected to both scale as area-law at leading-order with respect to the separating boundary~\cite{BELFIGLIO2024138398,1997-Mukohyama.etal-Phys.Rev.D,2020Chandran.ShankaranarayananPhys.Rev.D}. Since this is also the regime where Hayward and Bardeen black holes can be described by a de Sitter core, we expect the same behaviour for regular black holes close to $r=0$. To investigate this further, we study entanglement energy for the smallest possible subregion of the field corresponding to a sphere of radius $r_0=r[\rho=(N-1)a]$ in Fig. \ref{fig:Core}, upon varying core radius $r_c$. We may infer from Fig. \ref{fig:dScore} and \ref{fig:Core} that:
\begin{equation}
    S\sim c_s\frac{r^2}{a^2},\quad E \sim \frac{c_e r^2}{a^3},
\end{equation}
where the entanglement entropy continues to obey an area-law inside regular black-hole cores, and entanglement energy follows an area-law behaviour only as $r\to 0$ resembling its scaling for the Minkowski-flat background ($f(r)\to1$).

\subsection{Singular core}

\begin{figure*}[!ht]
	\begin{center}
		\subfloat[\label{dsc2d}][]{%
			\includegraphics[width=0.4\textwidth]{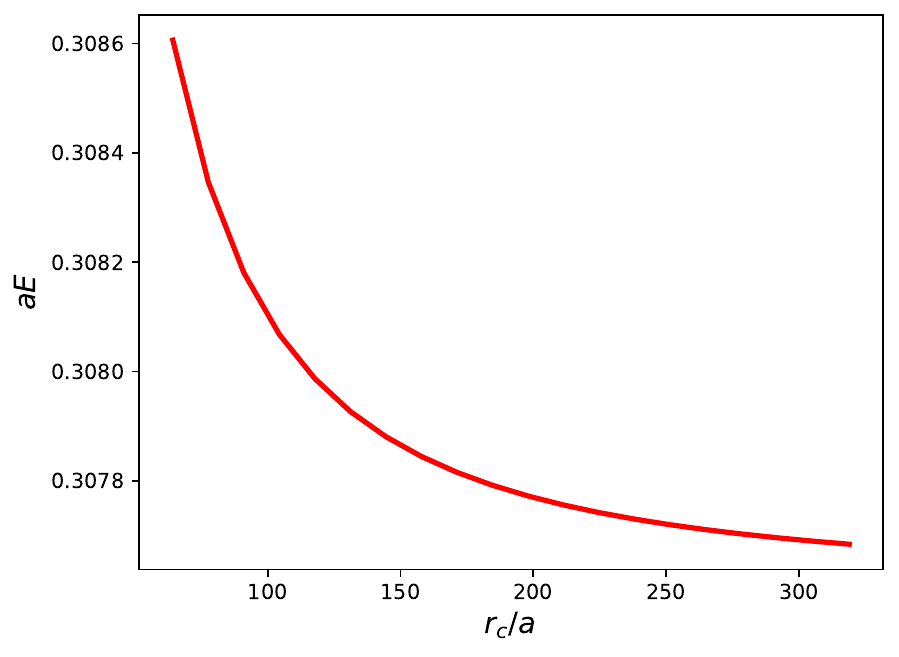}
		}
		\subfloat[\label{rnc2d}][]{%
			\includegraphics[width=0.4\textwidth]{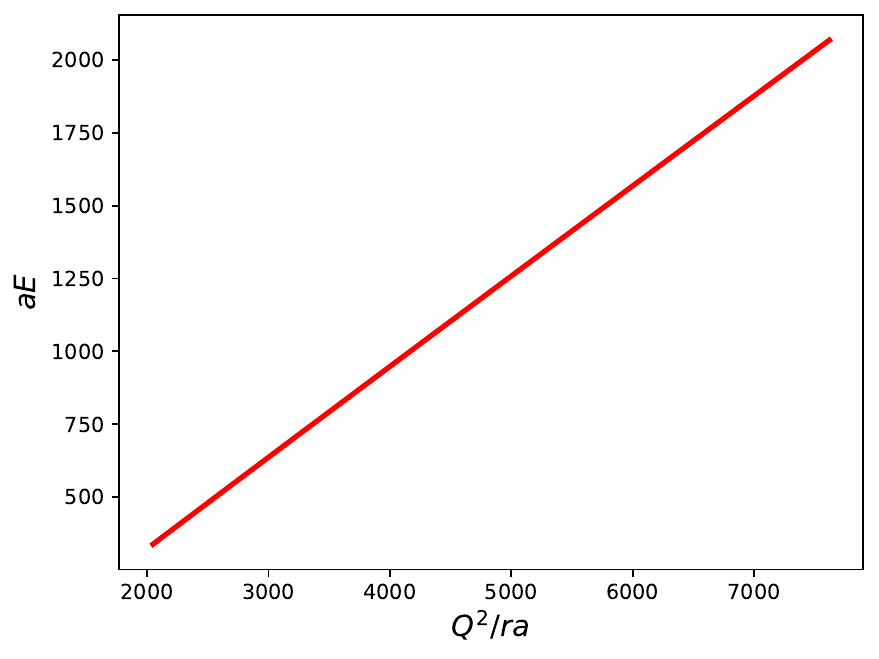}
		}
		
		\caption{Entanglement energy scaling for the smallest possible field subregion corresponding to a boundary at $\rho=(N-1)a$ from the inner horizon, or $r_0=r[\rho=(N-1)a]$ from the center. The plots are for (a) the de Sitter core of radius $r_c$ corresponding to regular black holes, and (b) the singular core of Reissner-Nordstr\"om (RN) with charge $Q$.}
		\label{fig:Core}
	\end{center}
\end{figure*}

Alternatively, let us here consider the case where the spacetime is singular as $r\to 0$, namely the case of standard black holes. Here, we restrict our attention to the Reissner-Nordstr\"om metric, describing a black hole with electric charge $Q$ and mass $M$:
\begin{equation}
    f(r)=1-\frac{2M}{r}+\frac{Q^2}{r^2}
\end{equation}
The proper length from the inner horizon is obtained as follows~\cite{2020Chandran.ShankaranarayananPhys.Rev.D}:
\begin{multline}
\rho=M\ln \left[\frac{\sqrt{M^2-Q^2}}{M-2\left\{r+\sqrt{r(r-2M)+Q^2}\right\}}\right]\\-\sqrt{r(r-2M)+Q^2}
\end{multline}
The IR cutoff is therefore fixed corresponding to $r=0$ as:
\begin{equation}
    \rho_L=(N+1)a=M \ln \left(\sqrt{\frac{M+Q}{M-Q}}\right)-Q
\end{equation}
From Fig. \ref{fig:RNBHCore}, we see that the entanglement entropy satisfies an area law throughout, similar to what we observed for the de Sitter core. However, the entanglement energy does not seem to converge to zero as $r\to0$, differently from the de Sitter case. Since $f(r)\sim Q^2/r^2$ close to the singularity, we investigate this further in Fig. \ref{fig:Core} by studying entanglement energy for the smallest possible subregion of the field confined to a sphere of radius $r_0=r[\rho=(N-1)a]$ for each value of charge $Q$.

From Fig. \ref{fig:Core}, we observe that the entanglement energy is largely constant for regular black holes ($aE\sim 0.31$), since it follows an area law ($\propto r_0^2)$ in this regime, and there is negligible change in $r_0$ with core radius $r_c$ (as per Fig. \ref{fig:dScore}). However, since $r_0$ is also found to change considerably with charge $Q$ in RNBH (as per Fig. \ref{fig:RNBHCore}), we are able to show precisely that the entanglement energy scales as electrostatic energy ($\propto Q^2/r_0$), thereby acting as a distinguisher between regular and singular spacetimes.

From Figs. \ref{fig:RNBHCore} and \ref{fig:Core}, we may infer that close to $r\sim 0$ RNBHs satisfy:
\begin{equation}
    S\sim c_s\frac{r^2}{a^2} \quad;\quad E \propto \frac{Q^2}{ra^2}
\end{equation}
where, in analogy with the de Sitter core, the entanglement entropy continues to follow an area law. However, the entanglement energy is found to be dominated by the electrostatic energy corresponding to a sphere of radius $r$ with charge $Q$, in direct contrast with the results for the de Sitter core.

This can be interpreted as a peculiar feature of singular cores, with no counterpart in regular scenarios such as Bardeen and Hayward black holes. A plausible explanation for this lies in the sensitivity of entanglement energy to the background. For instance, we have already established in Sec. \ref{sec:nearhorizon} that the entanglement energy scaling shifts from an area-law typical of Minkowski backgrounds (where $f(r)=1$)~\cite{1997-Mukohyama.etal-Phys.Rev.D}, to a linear scaling with Komar energy near the black hole horizon (where $f(r)\to0$)~\cite{2020Chandran.ShankaranarayananPhys.Rev.D}. On approaching the RNBH singularity we have $f(r)\sim Q^2r^{-2}\to\infty$, thereby giving rise to a new kind of scaling behaviour for entanglement energy, wherein it is proportional to the electrostatic energy. Along these lines, it was also suggested in \cite{1998-Mukohyama.etal-Phys.Rev.D} that this may be related to the gravitational redshifting of field modes due to the background curvature, which leaves an imprint on entanglement energy of the field but not on entanglement entropy, as the latter is invariant under such scaling transformations in quadratic Hamiltonians~\cite{2019-Chandran.Shankaranarayanan-Phys.Rev.D,2020Chandran.ShankaranarayananPhys.Rev.D}.

\section{Discussion and outlook}\label{sec:conc}

The entanglement features of a quantum field is capable of providing valuable insight into the thermodynamic properties of the background geometry. The ground state entanglement entropy, for instance, has been shown to reproduce the Smarr formula in certain static, spherically symmetric spacetimes~\cite{2020Chandran.ShankaranarayananPhys.Rev.D}. For regular black holes, such an approach offers a fresh perspective towards studying their thermodynamic properties, including their thermodynamic stability.

In Section \ref{sec:warmup}, we reviewed the standard discretization procedure for scalar fields in static, spherically symmetric configurations, deriving the coupling matrix for the field degrees of freedom on a lattice of spherical shells.

In Section \ref{sec:setup}, we argued for the universality of this approach by invoking the near-horizon approximation of the field degrees of freedom. We observed that the near-horizon Hamiltonian is completely parameterized by the Hawking temperature $T_H$ and the Bekenstein-Hawking entropy $S_{BH}$, regardless of the overall number of spacetime parameters involved. We also motivated the calculation of subsystem counterparts to thermodynamic measures, namely, entanglement entropy and entanglement energy, in order to rigorously study the near-horizon degrees of freedom.

In Section \ref{sec:nearhorizon}, we simulated the near-horizon entanglement entropy and entanglement energy for Bardeen and Hayward black holes. We showed that in the leading order, the entanglement entropy scaled proportionally to Bekenstein-Hawking entropy, whereas entanglement energy scaled proportionally to the Komar energy. From these scaling relations, we recovered the Smarr formula, i.e., the thermodynamic equation of state describing each of the regular black holes in question. We also showed that both black holes underwent a second-order phase transition wherein the heat capacity diverged and flipped its sign from negative (thermodynamically unstable) to positive (thermodynamically stable) upon approaching the extremal limit.

In Section \ref{sec:core}, we simulated the entanglement entropy and entanglement energy of the field degrees of freedom inside the core to obtain distinct signatures of regularity. While entanglement entropy exhibited a typical area-law for both regular and singular cores, entanglement energy assumed different scaling behaviours for singular and regular black holes. Upon further investigating the smallest field subregion about $r=0$, we inferred that for the de-Sitter core (corresponding to regular black holes), the entanglement energy followed an area law close to $r\to0$, whereas for the singular core of Reissner-Nordstr\"om black holes, it exhibited a scaling proportional to electrostatic energy as $E\propto Q^2/r$. This implied that, while the near-horizon results established a universal connection between entanglement energy and Komar energy, our outcomes near $r=0$ clearly discriminated between regular and singular solutions, showing sensitivity of the entanglement energy to the background.

Looking ahead, we plan to study the correspondence between entanglement and black hole thermodynamics in some non-static configurations (e.g. rotating black holes), in order to understand whether the issue of black hole stability can be addressed from entanglement considerations in more general scenarios. Furthermore, the presence of non-minimal coupling between quantum fields and the background geometry may contribute to non-trivial effects that require further investigation. Finally, it would be interesting to quantify possible deviations associated with quantum effects at Planck length scales, where the classical description of black holes is no longer sufficient. In the absence of a complete theory of quantum gravity, some effective frameworks have been proposed, suggesting deviations from classical results at sufficiently small scales. Such quantum effects may alter black hole and entanglement thermodynamics near the core, thus allowing the isolation of possible ``quantum" signatures in black hole scenarios.

\section*{Acknowledgements}
SMC is supported by Prime Minister's Research Fellowship offered by the Ministry of Education, Govt. of India. S.M. acknowledges support from Italian Ministry of Universities and Research under “PNRR MUR project PE0000023-NQSTI”.

\appendix

\section*{Appendix A: Notes on black hole thermodynamics}

In this appendix, we recap some useful steps on how to construct the black hole thermodynamics we employed in the main text.

Precisely, information on the gravitational charge and thermodynamics of black holes is fully-contained inside the first principle, where an appropriate effective mass, $m_{{\rm eff}}$, plays the role of gravitational charge.

While in the Schwarzschild solution we simply have $M=m_{{\rm eff}}$,  finding the role of such effective mass is generally not trivial.

In analogy with standard thermodynamics, it appears natural to compute the heat capacity at constant pressure by starting from the Smarr formula of Eq. \eqref{eq:Smarr}, having
\begin{equation}\label{standardC}
    \mathcal C=\frac{\partial h}{\partial T}=\frac{\partial m_{{\rm eff}}}{\partial T}=T\frac{\partial S_{BH}}{\partial T}\,,
\end{equation}
where $h$ is the black hole enthalpy.

In general, the above formula can acquire negative values \cite{Davies:1977bgr}, albeit possible solutions to this problem are under consideration \cite{DAgostino:2024ymo}.

However, in our treatment the gravitational charge might be the Komar mass, namely $m_{eff}=E_{Komar}$, satisfying $m_{eff}=2T_HS_{BH}$, when evaluated at the horizon~\cite{2010-Banerjee.Majhi-Phys.Rev.D,2010BanerjeePhys.Rev.D}.

If we compare this property with the case of Reissner-Nordstr\"om black hole, we have
\begin{equation}
    m_{eff}=M-\frac{Q^2}{r_h}=2T_HS_{BH}\implies M=2T_HS_{BH}+\underbrace{\left(\frac{Q}{r_h}\right)}_{\Phi_H}Q,
\end{equation}
observing that, even if $m_{eff}\neq M$, we can write down a thermodynamic equation of state relating $M$, $T_H$, $S_{BH}$ and other parameters.

Indeed, if we compare $
    dm_{eff}=\frac{M}{2S_{BH}}dS_{BH}-\Phi_HdQ$ and $dM=T_HdS_{BH}+\Phi_HdQ$, we then observe an inconsistency in the former case: In fact, from here it is clear that the first law in terms of $m_{eff}$ does not give the temperature of Reissner-Nordstr\"om black hole as opposed to the first law in terms of $M$. For the same reason, it makes sense to define the heat capacity from the first law  written in terms of $dM$ instead of $dm_{eff}$, as followed by Davies~\cite{1978-Davies-ReportsonProgressinPhysics,1977HutMonthlyNoticesoftheRoyalAstronomicalSociety}.

\end{document}